\documentclass[showpacs,preprintnumbers,amsmath,amssymb,twocolumn]{revtex4}
\usepackage{amsmath,amsfonts,latexsym,amssymb,graphicx,graphics,epsfig,subfigure,color,makeidx}
\usepackage{xcolor,diagbox}
\usepackage{multirow}
\usepackage[colorlinks,linkcolor=blue,anchorcolor=blue,citecolor=green,urlcolor=blue]{hyperref}

\usepackage[latin1]{inputenc}

\newcommand {\nn}    {\nonumber}
\newcommand {\fc}    {\frac}
\newcommand {\be}    {\begin{equation}}
\newcommand {\ee}    {\end{equation}}
\newcommand {\beq}   {\begin{eqnarray}}
\newcommand {\eeq}   {\end{eqnarray}}
\newcommand {\pd}    {\partial}
\newcommand {\lt}    {\left}
\newcommand {\rt}    {\right}

\begin{document}

\title{Quasinormal modes of a charged black hole with scalar hair}

\author{Wen-Di Guo$^{ab}$\footnote{guowd@lzu.edu.cn,Wen-Di Guo and Qin Tan are co-first authors of this paper.}}
\author{Qin Tan$^{ab}$\footnote{tanq19@lzu.edu.cn}}

\affiliation{$^{a}$Lanzhou Center for Theoretical Physics, Key Laboratory of Theoretical Physics of Gansu Province,
School of Physical Science and Technology, Lanzhou University, Lanzhou 730000, People's Republic of China\\
             $^{b}$Institute of Theoretical Physics $\&$ Research Center of Gravitation, Lanzhou University, Lanzhou 730000, People's Republic of China}

\begin{abstract}
From a five-dimensional Einstein-Maxwell theory, Bah et al. constructed a singularity free topology star/black hole [Phys. Rev. Lett. 126, 151101 (2021)]. After the Klein-Kluza reduction, i.e., integrating the extra space dimension, it can obtain an effective four-dimensional static spherical charged black hole with scalar hair. In this paper, we study the quasinormal modes (QNMs) of the scalar field, electromagnetic field, and gravitational field on the background of this effective four-dimensional charged black hole. The radial parts of the perturbed fields all satisfy a Schr\"{o}dinger-like equation. Using the asymptotic iteration method, we obtain the QNM frequencies semianalytically. For low overtone QNMs, the results obtained from the asymptotic iteration method and the Wentzel-Kramers-Brillouin approximation method agree well. In the null coordinates, the evolution of a Gaussian package is also studied. The QNM frequencies obtained by fitting the evolution data also agree well with the results obtained by the asymptotic iteration method.
\end{abstract}



\maketitle

\section{Introduction}

In 2015, Laser Interferometer Gravitational-Wave Observatory (LIGO) and Virgo detected the gravitational wave of a binary black hole system~\cite{Collaboration2016}. This opens the window to gravitational wave astrophysics. After that, the Event Horizon Telescope (EHT) took the first picture of the supermassive black hole at the center of galaxy M87 in 2019~\cite{Collaboration2019,Collaboration2019a,Collaboration2019b,Collaboration2019c,Collaboration2019d,Collaboration2019e} and the picture of the black hole in our Milky Way in 2022~\cite{Collaboration2022,Collaboration2022a,Collaboration2022b,Collaboration2022c,Collaboration2022d,Collaboration2022e}. These enhanced our ability of testing fundamental physical problems. One of them is that does singularity exist~\cite{Berti2015,Cardoso2016}? Searching for black hole alternatives has been attracted a lot of interest. Ultra-compact objects such as gravastars~\cite{Mazur2001}, boson stars~\cite{Schunck2008}, wormholes~\cite{Solodukhin2005,Dai:2019mse,Simonetti:2020ivl,Bambi:2021qfo} have been constructed. More details can be seen in the review~\cite{Cardoso2019} and references therein. However, these models usually need some exotic matters and their UV origin is not clear. On the other hand, from string theory, the most important candidate which can unify quantum theory and gravity theory, some horizonless models have been constructed. One of them, fuzz ball, has smooth microstate geometry and it is similar to classical black hole up to Planck scale~\cite{Gibbons2013}. But constructing fuzz ball needs a lot of degrees of freedom, and it is difficult to study the astrophysical observations~\cite{Bena2020,Bena2020a,Bena2020b}. In order to alleviate these disadvantages, Bah and Heidmann proposed a topological star/black hole model which can be constructed from type-II B string theory and it is similar to classical black holes on macrostate geometries~\cite{Bah2020,Bah2020a}. So studying the astrophysical observations is not so difficult. They further analysed the thermodynamic stability carefully of this solution~\cite{Bah:2021irr}. Besides, the motion of a charged particle on the background of this topological star/black hole model has been studied~\cite{Lim2021}. Based on this solution, a four-dimensional static spherical charged black hole with scalar hair can be obtained by integrating the extra dimension~\cite{Bah2020,Bah2020a}. We will study the quasinormal modes (QNMs) of the scalar field, electromagnetic field and gravitational field on the background of this black hole in this paper.

Quasinormal modes are the characteristic modes of dissipative systems and have been playing important roles in a lot of physical areas. Classically, everything falling into the interior of the black hole can not escape, because of the event horizon, so black holes are dissipative systems. Quasinormal modes of black holes dominates the ringdown stage, the final stage of gravitational waves for a binary black hole merging system~\cite{Berti2007}. A big difference between the QNMs and normal modes is that the eigenfunctions of normal modes form a complete set, but the QNMs do not~\cite{Nollert1998}. Besides, the eigenfunctions of QNMs are not normalizable~\cite{Nollert1998}. Quasinormal modes have complex frequencies,
the real parts are the vibration frequencies, and the imaginary parts are the inverse of the decay time scale of the perturbation. It is very important to study the QNMs of black holes. The mass and angular momentum can be inferred from the QNMs, and the no-hair theorem can also be tested through QNMs~\cite{Berti:2005ys,Berti:2007zu,Isi:2019aib}. For horizonless compact objects, there could be echoes in the ringdown signal, which is the smoking gun of the existence of the horizonless compact objects~\cite{Cardoso2019,Cardoso:2016rao,Cardoso:2017cqb}. One can also use the QNMs to constrain modified gravity theories~\cite{Wang:2004bv,Blazquez-Salcedo:2016enn,Franciolini:2018uyq,Aragon:2020xtm,Liu:2020qia,Karakasis:2021tqx,Cano:2021myl,Gonzalez:2022upu,Zhao:2022gxl}.
It is also found the QNM spectrum is unstable under the small perturbation of the potential~\cite{Jaramillo:2020tuu,Cheung:2021bol}. Besides, the QNM frequencies can also partly reveal the stability of the background spacetime under small perturbations~\cite{Ishibashi:2003ap,Chowdhury:2022zqg}. In other physical systems, QNMs also play very important roles, for example, leaky resonant cavities~\cite{Kristensen2015}, and brane world models~\cite{Seahra2005,Seahra2005a,Tan2022}. It has been studied widely~\cite{Cai:2015fia,Cardoso:2019mqo,McManus:2019ulj,Cardoso:2020nst,Guo:2021enm,Hatsuda:2020sbn,Noda:2022zgk}.

In this paper, we will study the QNMs on the background of the four-dimensional static spherical black hole with scalar hair. The paper is organized as follows. In Sec.~\ref{themodel}, we give a brief review on the charged black hole with scalar hair and the KK reduction. In Sec.~\ref{pereqs}, we study the small perturbation of scalar field, electromagnetic field and gravitational field. Expanding the perturbation fields in spherical harmonic function, we can derive the master equations of the perturbation. In Sec.~\ref{QNM}, we compute the QNM frequencies using the asymptotic iteration method (AIM) and Wentzel-Kramers-Brillouin (WKB) approximation method, and the time evolution of a Gaussian package. Finally, we give the conclusions in Sec.~\ref{conclusion}.

\section{The charged black hole}\label{themodel}

The nonsingular black hole/topology star proposed by I. Bah and P. Hedmann~\cite{Bah2020,Bah2020a} started from the action of a five-dimensional Einstein-Maxwell theory
\begin{equation}
S=\int d^5x\sqrt{-\hat{g}}\left(\fc{1}{2\kappa_5^2}\hat{R}-\fc{1}{4}\hat{F}^{MN}\hat{F}_{MN}\right),
\end{equation}
where $\hat{F}_{MN}$ is the five-dimensional electromagnetic field tensor and $\kappa_{5}$ is the five-dimensional gravitational constant. We use the hat to denote the five-dimensional quantities. And the capital Latin letters $M, N...$ are used to denote the five-dimensional coordinates. They considered a spherically symmetric metric ~\cite{Stotyn2011}
\begin{eqnarray}
ds^2&=&-f_S(r)dt^2+f_B(r)dy^2+\frac{1}{f_S(r)f_B(r)}dr^2\nn\\
&+&r^2d\theta^2+r^2\sin^2\theta d\phi^2. \label{metric_five}
\end{eqnarray}
The extra dimension coordinate is denoted by $y$. The field strength of the magnetic field is
\be
\hat{F}=P\sin\theta d\theta\wedge d\phi.\label{field_strength}
\ee
The solution can be solved as~\cite{Stotyn2011}
\be
f_{B}(r)=1-\frac{r_B}{r},  ~~~~~~~ f_{S}(r)=1-\frac{r_S}{r}, ~~~~~~~  P=\pm\frac{1}{\kappa_{5}^{2}}\sqrt{\frac{3r_Sr_B}{2}}\label{solution}.
\ee
Metric~\eqref{metric_five} is symmetric under the double rotation, which transforms the coordinate ($t$, $y$, $r_S$, $r_B$) to ($iy$, $it$, $r_B$, $r_S$). The spacetime at $r=r_S$ is a horizon, where $f_{S}(r)=0$, and the spacetime at $r=r_B$ is a degeneracy of the $y$-circle. Bah and Heidmann showed that there are smooth bubbles at $r=r_B$ which ends the spacetime~\cite{Bah2020,Bah2020a}. The spacetime has two configurations. One is black string, when $r_S\geq r_{B}$, because the bubble hides behind the horizon. The other is topology star, when $r_S<r_{B}$, the horizon disappear because the spacetime ends at $r=r_B$~\cite{Bah2020,Bah2020a}.

We rewrite the metric~\eqref{metric_five} as
\beq
ds_5^2&=&e^{2\Phi}ds_4^2+e^{-4\Phi}dy^2,\\
ds_4^2&=&
         f_B^{\frac{1}{2}}\lt(-f_Sdt^2+\frac{dr^2}{f_Bf_S}+r^2d\theta^2+r^2\sin^2\theta d\phi^2\rt)\label{metric_four},
\eeq
where
\be
e^{2\Phi}=f_B^{-1/2},
\ee
and $\Phi$ is a dilaton field. After the Kaluza-Klein reduction, i.e., integrating the extra dimension, we can obtain a four-dimensional Einstein-Maxwell-dilaton theory
\beq
S_4&=&\int d^4x\sqrt{-g}\Big(\fc{1}{2\kappa_4^2}R_4-\fc{3}{\kappa_4^2}g^{\mu\nu}\pd_{\mu}\Phi\pd_{\nu}\Phi\nn\\
&-&2\pi R_y e^{-2\Phi}F_{\mu\nu}F^{\mu\nu}\Big),
\eeq
where $R_y$ is the radius of the extra dimension. We use the Greek letters $\mu, \nu...$ to label the four-dimensional coordinates. Here, the quantities without hat are constructed in four-dimensional spacetime. The four-dimensional gravitational constant is defined as
\be
 \kappa_4=\fc{\kappa_5}{\sqrt{2\pi R_y}}.
\ee
We can solve the four-dimensional field strength of the magnetic field as
\be
F=\pm\fc{1}{\kappa_4\sqrt{2\pi R_y}}\sqrt{\fc{3r_B r_S}{2}}\sin\theta d\theta\wedge d\phi.
\ee
The ADM mass $M$ and the magnetic charge $Q_\text{m}$ can be solved as
\beq
M&=&2\pi\left(\fc{2r_S+r_B}{\kappa_4^2}\right),\nn\\
Q_\text{m}&=&\fc{1}{\kappa_4}\sqrt{\fc{3}{2}r_B r_S}.
\eeq
Or, in terms of $M$ and $Q_\text{m}$
\beq
r_S^{(1)}&=&\fc{\kappa_4^2}{8\pi}(M-M_{\triangle}),~~~~~r_B^{(1)}=\fc{\kappa_4^2}{4\pi}(M+M_{\triangle}),\label{MQ1}\\
r_S^{(2)}&=&\fc{\kappa_4^2}{8\pi}(M+M_{\triangle}),~~~~~r_B^{(2)}=\fc{\kappa_4^2}{4\pi}(M-M_{\triangle}),\label{MQ2}
\eeq
where $M_{\triangle}^2=M^2-\left(\fc{8\pi Q_\text{m}}{\sqrt{3}\kappa_4}\right)^2$. From solution~\eqref{solution} we know that, when $r<r_B$, $f_B^{1/2}$ becomes imaginary, and the metric is unphysical. So, the coordinate $r$ can not smaller than $r_B$. Besides, the the black string scenario has Gregory-Laflamme instability~\cite{Gregory:1993vy}. However, the models with compact extra dimensions which will lead to a discrete KK mass spectrum could avoid the Gregory-Laflamme instability. Stotyn and Mann have demonstrated that, when $R_y>\fc{4\sqrt{3}}{3}Q_{\text{m}}$, solution~\eqref{MQ2} is stable. Note that, the spacetime at $r=r_B$ is singular, so when $r_S\geq r_B$ the metric~\eqref{metric_four} describes a charged black hole with scalar hair. We only study this case in this paper. Note that, the magnetically charged black holes have been studied in Refs.~\cite{Ghosh:2020tdu,Diamond:2021scl,Karas:2023uac}.

\section{Perturbation equations}\label{pereqs}

In this section, we analyse the linear perturbation equations of the scalar field, electromagnetic field, and gravitational field on the background of the charged black hole with scalar hair. For simplicity, we consider the three kinds of perturbed fields separately, that is to say, when one field is perturbed, the background of the other two fields are not affected. Although, usually the perturbations of the electromagnetic field and gravitational field should be coupled together for charged black holes.

\subsection{Scalar field}
We consider a free massless scalar field on this charged black hole background. The equation of motion for the scalar field is the Klein-Gordon equation
\be
\fc{1}{\sqrt{-g}}\pd_{\mu}\lt({\sqrt{-g}\pd^{\mu}\varphi}\rt)=0\label{eqphi}.
\ee
Because of the spherical symmetry and time independence of the background, we can decompose the scalar field as
\be
\varphi(t,r,\theta,\phi)=\sum_{l,m}e^{-i\omega t}f_B^{-1/4}\fc{1}{r}\psi_{\text{s}}(r)Y_{l,m}(\theta,\phi),
\ee
where $Y_{l,m}$ is the spherical harmonics which satisfies
\be
\triangle Y_{l,m}=-l(l+1)Y_{l,m},
\ee
with $\triangle$ the Laplace-Beltrami operator. Substituting this into Eq.~\eqref{eqphi}, we obtain the radial part of the  perturbation equation for the scalar field
\be
f_S^2f_B \psi_{\text{s}}''+f_S(f_S'f_B+\fc{1}{2}f_B'f_S)\psi_{\text{s}}'+(\omega^2-V_{\text{s}}(r))\psi_{\text{s}}=0,\label{speq}
\ee
where
\beq
V_{\text{s}}(r)&=&f_S\lt(\fc{l(l+1)}{r^2}+\fc{1}{4}f_Sf_B''+f_B'f_S'-\fc{f_Sf_B'^{2}}{4f_B}\rt)\nn\\
&+&\fc{f_S}{r}(f_Sf_B'+f_Bf_S')
\eeq
is the effective potential for the scalar field. Hereafter, we use prime to denote the derivative with respect to the coordinate $r$. In order to obtain the Schr\"{o}dinger-like equation, we need the tortoise coordinate $r_*$, which can be obtained from the following relation
\be
dr_*=\fc{1}{\sqrt{f_B}f_S}dr\label{tort}.
\ee
In this way, Eq.~\eqref{speq} can be written as
\be
\fc{d^2\psi(r_*)}{dr_*^2}+(\omega^2-V_{\text{s}}(r_*))\psi(r_*)=0\label{schoreq}.
\ee
\subsection{Electromagnetic field}
For an electromagnetic field, the Maxwell equation is given by
\be
\fc{1}{\sqrt{-g}}\pd_{\mu}\lt(\sqrt{-g}\tilde{F}^{\mu\nu}\rt)=0,\label{eqAu}
\ee
where $\tilde{F}_{\mu\nu}=\pd_\mu \tilde{A}_{\nu}-\pd_\nu \tilde{A}_{\mu}$ is the field strength tensor of the perturbed electromagnetic field $\tilde{A_{\mu}}$. To separate the perturbed electromagnetic field, we need the vectorial spherical harmonics which are defined as~\cite{Wheeler,Ruffini,Edmonds}
\beq
(V^1_{l,m})_a&=&\pd_a Y_{l,m}(\theta,\phi), \\
(V^2_{l,m})_a&=&\gamma^{bc}\epsilon_{ac}\pd_b Y_{l,m}(\theta,\phi).
\eeq
Here, the Latin letters $a, b, c$ denote the angular coordinates $\theta$ and $\phi$, $\gamma$ is the induced metric on the sphere with radius $1$, and $\epsilon$ is the totally antisymmetric tensor in two dimensions. Note that, $(V^1_{l,m})_a$ and $(V^2_{l,m})_a$ behave different under the space inversion, i.e., $(\theta,\phi)\rightarrow(\pi-\theta,\pi+\phi)$. $(V^1_{l,m})_a$ is even or polar, that is, it acquires a factor $(-1)^l$ under space inversion, $(V^2_{l,m})_a$ is odd or axial, that is, it acquires a factor $(-1)^{l+1}$ under space inversion. Thus, the perturbed electromagnetic field $\tilde{A}_{\mu}$ can be decomposed as
\beq
\tilde{A}_{\mu}(t,r,\theta,\phi)&=&\sum_{l,m}e^{-i\omega t}\lt[\begin{array}{c}
0\\
0\\
\fc{\psi_{\text v}(r)}{\sin\theta}\fc{\pd Y_{l,m}}{\pd\phi}\\
-\psi_{\text v}(r)\sin\theta\fc{\pd Y_{l,m}}{\pd\theta}
\end{array} \rt]\nn\\
&+&\sum_{l,m}e^{-i\omega t}\lt[\begin{array}{c}
h_1(r)Y_{l,,m}\\
h_2(r)Y_{l,,m}\\
h_3(r)\fc{\pd Y_{l,m}}{\pd\theta}\\
h_3(r)\fc{\pd Y_{l,m}}{\pd\phi}
\end{array} \rt].
\eeq

Owing to the spherical symmetry of the background metric, the perturbation equations will not mix polar and axial contributions. Besides, the axial part and the polar part will contribute to the same result~\cite{Wheeler,Ruffini}. So, we only need to deal with the axial part. Substituting the background metric~\eqref{metric_four} into the Maxwell equation~\eqref{eqAu}, we obtain the perturbation equation for the radial part $\psi_{\text{v}}$
\be
f_S^2f_B \psi_{\text{v}}''+f_S(f_S'f_B+\fc{1}{2}f_B'f_S)\psi_{\text{v}}'+(\omega^2-V_{\text{v}}(r))\psi_{\text{v}}=0,\label{vpeq}
\ee
where the effective potential is
\be
V_{\text{v}}(r)=\fc{f_S(r)l(l+1)}{r^2},
\ee
which does not depend on the parameter $r_B$. But the effective potential depends on the parameter $r_S$ which is related to the magnetic charge $Q_{\text{m}}$. Using the tortoise coordinate $r_*$, the perturbation equation can also be transformed into a Schr\"{o}dinger-like form. From Eq.~\eqref{tort} we know that, the tortoise coordinate $r_*$ depends on the parameter $r_B$, so the QNMs will also be affected by the parameter $r_B$.
\subsection{Gravitational field}
Considering a perturbation on the background metric~\eqref{metric_four}, the perturbed metric is
\be
\bar{g}_{\mu\nu}=g_{\mu\nu}+h_{\mu\nu},
\ee
where $h_{\mu\nu}$ is the perturbation. The separation of the perturbation for the gravitational field is more complicated. Besides the vectorial spherical harmonics, we also need the tensorial harmonics, which are defined as~\cite{Regge}
\beq
(T^1_{l,m})_{ab}&=&(Y_{l,m});_{ab}, \\
(T^2_{l,m})_{ab}&=&Y_{l,m}\gamma_{ab},\\
(T^3_{l,m})_{ab}&=&\fc{1}{2}\lt[\epsilon_a^c(Y_{l,m});_{cb}+\epsilon_b^c(Y_{l,m});_{ca}\rt],
\eeq
where the semicolon denotes the covariant derivative on the sphere. Among them, $T^3_{l,m}$ is odd under the space inversion, the other two are even. Based on the principle of general covariance, the theory should keep covariant under infinitesimal coordinate transformation. Thus, we can choose a specific  gauge to simplify the problem. In the Regge-Wheeler gauge~\cite{Regge}, the perturbation $h_{\mu\nu}$ can be written as
\be
h_{\mu\nu}=\sum_{l}e^{-i\omega t}\lt[\begin{array}{cccc}
0&0&0&h_0(r)\\
0&0&0&h_1(r)\\
0&0&0&0\\
h_0(r)&h_1(r)&0&0
\end{array}\rt]\sin\theta\pd_{\theta}Y_{l,0}(\theta)
\ee
for the odd parity, and
\be
h_{\mu\nu}=\sum_{l}e^{-i\omega t}\lt[\begin{array}{cccc}
H_{0}(r)&H_{1}(r)&0&0\\
H_{1}(r)&H_{2}(r)&0&0\\
0&0&r^2K(r)&0\\
0&0&0&r^2K(r)\sin^2\theta
\end{array}\rt]Y_{l,0}(\theta)
\ee
for the even parity. Note that, we have chosen $m=0$ for simplicity, because the perturbation equations do not depend on the value of $m$~\cite{Regge}. For the Schwarzschild black hole, the odd parity and the even parity have the same QNM spectrum ~\cite{Chandrasekhar}, but other black holes may not exist this property. Anyway, for simplicity, we just study the odd parity in this paper. Substituting the decomposition into Einstein's equation, and after some algebraic operations, the perturbation equations for the odd parity perturbation can be combined into a single equation for the variable $\psi_{\text{g}}$ , which is defined as
\be
\psi_{\text{g}}(r)=\fc{f_B^{1/4}(r)f_S(r)}{r}h_1(r).
\ee
The variable $\psi_{\text{g}}$ satisfies the Schr\"{o}dinger-like equation~\eqref{schoreq}, with the effective potential
\beq
V_{\text{g}}(r)&=&f_S \left(\frac{3}{4} f_B' f_S'+f_B f_S''-\frac{f_B f_S'}{r}+\frac{l(l+1)}{r^2}\right)\nn\\
&-&f_S^2 \left(\frac{9 f_B'^2}{16 f_B}-\frac{3 f_B''}{4}\right).
\eeq

\begin{figure*}[htb]
\begin{center}
\subfigure[$ $]  {\label{VS}
\includegraphics[width=5.7cm]{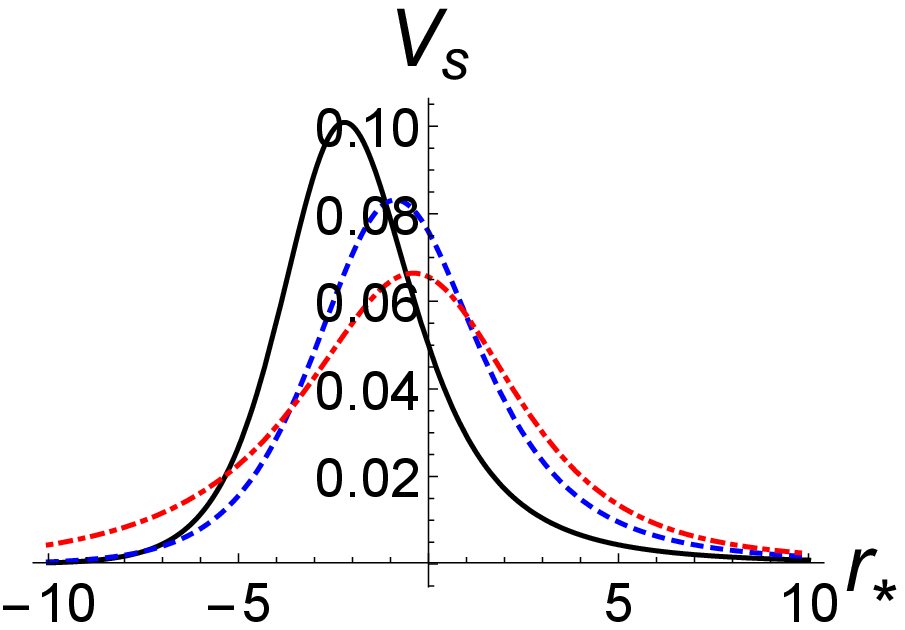}}
\subfigure[$ $]  {\label{VV}
\includegraphics[width=5.7cm]{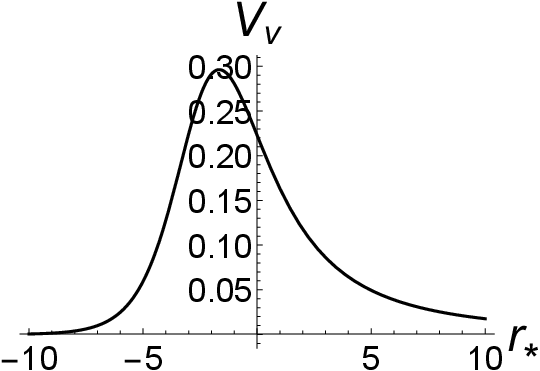}}
\subfigure[$ $]  {\label{VG}
\includegraphics[width=5.7cm]{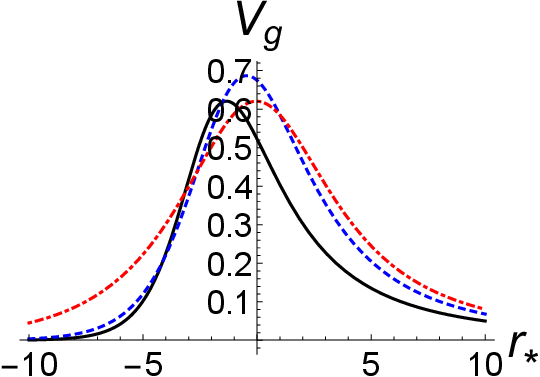}}
\end{center}
\caption{The effective potentials in the tortoise coordinate $r_*$. The parameter $r_B$ is set to $r_B=0.1r_S$ (the black solid lines), $r_B=0.5 r_S$ (the blue dashed lines), and $r_B=0.9 r_S$ (the red dot dashed lines). (a) The scalar field with $l=0$. (b) The electromagnetic field with $l=1$. (c) The gravitational field with $l=2$.}
\label{potential}
\end{figure*}

The effective potential plays an important role on determining the value of QNMs. We plot the effective potentials for scalar field, electromagnetic field, and gravitational field in Fig.~\ref{potential}. All of the effective potentials approach to zero at the horizon and infinity, which is similar to that of the Schwarzschild black hole. Besides, the effective potentials for the scalar field and the gravitational field depend on the parameter $r_B$.

\section{Computing QNMs}\label{QNM}

In this section we will solve the QNMs of the charged black hole with scalar field both in frequency domain and time domain. And we will compare the results by the two methods.

\subsection{Solving frequency}

First, we solve the frequencies of QNMs for the three kinds of fields using the AIM and WKB approximation method. Now, we give a brief review on the AIM.

\begin{figure*}[htb]
\begin{center}
\subfigure[$ $]  {\label{qnmsb05}
\includegraphics[width=5.7cm]{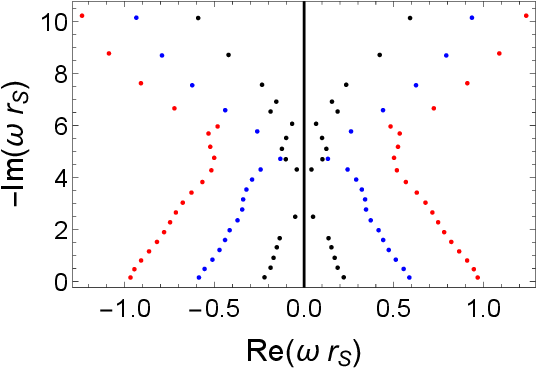}}
\subfigure[$ $]  {\label{qnmvb05}
\includegraphics[width=5.7cm]{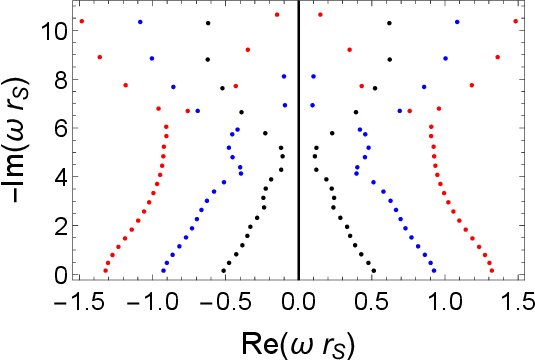}}
\subfigure[$ $]  {\label{qnmtb05}
\includegraphics[width=5.7cm]{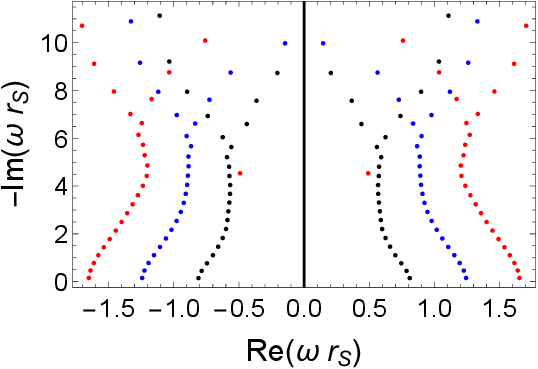}}
\end{center}
\caption{The first twenty QNMs for this charged black hole with scalar hair. The parameter $r_B$ is set to $r_B=0.5r_S$.  (a) QNMs for the scalar field with $l=0$ (the black dots), $l=1$ (the blue dots), $l=2$ (the red dots). (b)  QNMs for the electromagnetic field with $l=1$ (the black dots), $l=2$ (the blue dots), $l=3$ (the red dots). (c)  QNMs for the gravitational field with $l=2$ (the black dots), $l=3$ (the blue dots), $l=4$ (the red dots).}
\label{qnmsvt}
\end{figure*}

Considering  a  homogeneous linear second-order differential equation
\be
\chi''(x)=\lambda_0(x)\chi'(x)+s_0(x)\chi(x),\label{sdeq}
\ee
where $\lambda_0(x)$ and $s_0(x)$ are smooth functions. Based on the symmetric structure of the right-hand of Eq.~\eqref{sdeq}, we can find a general solution of this equation~\cite{Ciftci}. Differentiating Eq.~\eqref{sdeq} with respect to the variable $x$, we obtain
\be
\chi'''(x)=\lambda_1(x)\chi'(x)+s_1(x)\chi(x),\label{sdeq1}
\ee
where
\beq
\lambda_1(x)&=&\lambda_0'(x)+s_0(x)+(\lambda_0)^2,\nn\\
 s_1(x)&=&s_0'(x)+s_0(x)\lambda_0(x).
\eeq
Differentiating Eq.~\eqref{sdeq1} with respect to $x$ again, we find that
\be
\chi''''(x)=\lambda_2(x)\chi'(x)+s_2(x)\chi(x),\label{sdeq2}
\ee
where
\beq
\lambda_2(x)&=&\lambda_1'(x)+s_1(x)+\lambda_0\lambda_1, \nn\\
s_2(x)&=&s_0'(x)+s_0(x)\lambda_1(x).
\eeq
Continue this process, the $(n+1)-{th}$ and $(n+2)-{th}$ derivatives give us the following relations
\beq
\lambda_n(x)&=&\lambda_{n-1}'(x)+s_{n-1}(x)+\lambda_0(x)\lambda_{n-1}(x),\nn\\
 s_n(x)&=&s_{n-1}'(x)+s_0(x)\lambda_{n-1}(x).
\eeq
When $n$ is sufficiently large, the asymptotic aspect can be introduced as
\be
\fc{s_n(x)}{\lambda_n(x)}=\fc{s_{n-1}(x)}{\lambda_{n-1}(x)}.
\ee
Then, the QNMs can be obtained from the``quantization condition"
\be
s_n\lambda_{n-1}-s_{n-1}\lambda_n=0.
\ee
This is equivalent to give the iteration number $n$ a truncation. This method is good, but one needs to differentiate the $s(x)$ and $\lambda(x)$ terms for each iteration, which may bring problems for numerical precision. To make the process more efficient, Cho et al.~\cite{Cho} improved this method. The improved method does not need to take derivatives at each step, which greatly improves the accuracy and the speed. What they do is expanding the $\lambda_n$ and $s_n$ in a Taylor series around a fix point $\xi$,
\beq
\lambda_{n}(x)&=&\sum_{i=0}^{\infty}c_n^i(x-\xi)^i,\nn\\
s{n}(x)&=&\sum_{i=0}^{\infty}d_n^i(x-\xi)^i,
\eeq
where $c_n^i$ and $d_n^i$ are the $i-{th}$ Taylor coefficients of $\lambda_n$ and $s_n$, respectively. Using these expressions, one can obtain a set of recursion relations:
\beq
c_n^i&=&(i+1)c_{n-1}^{i+1}+d_{n-1}^i+\sum_{k=0}^{i}c_0^k c_{n-1}^{i-k},\nn\\
d_n^i&=&(i+1)d_{n-1}^{i+1}+\sum_{k=0}^{i}d_0^k c_{n-1}^{i-k}.
\eeq
Thus, the ``quantization condition" can be reexpressed in terms of the coefficients as
\be
d_n^0 c_{n-1}^0-d_{n-1}^0c_n^0=0.
\ee
The boundary conditions are pure ingoing waves at the event horizon
\be
\psi\sim e^{-i\omega r_*},~~r_*\rightarrow -\infty,
\ee
and pure outgoing waves at spatial infinity
\be
\psi\sim e^{i\omega r_*},~~r_*\rightarrow +\infty.
\ee
It is helpful to transform the infinity to be finite, so we perform the following coordinate transformation
\be
u=1-\fc{r_S}{r},
\ee
such that, the range of $u$ is $0\leq u<1$. The boundary conditions in terms of $u$ are
\be
\psi(u)=\lt(\fc{-2 (r_S-r_B)+2\sqrt{r_s-r_B}}{r_S^{5/2}u}\rt)^{i\omega r_S^{3/2}/\sqrt{r_S-r_B}}
\ee
at the horizon, i.e., $u=0$, and
\be
\psi(u)=e^{i\omega\lt(r_S/(1-u)+(r_S+r_B/2)\rt)}\lt(\fc{4r_S}{1-u}-r_B\rt)^{i\omega}
\ee
at infinity, i.e., $u\rightarrow 1$. Thus, we can define
\beq
\psi(u)&=&\lt(\fc{-2 (r_S-r_B)+2\sqrt{r_s-r_B}}{r_S^{5/2}u}\rt)^{i\omega r_S^{3/2}/\sqrt{r_S-r_B}}\nn\\
&\times&\lt(\fc{4r_S}{1-u}-r_B\rt)^{i\omega}e^{i\omega\lt(\fc{r_S}{(1-u)}+(r_S+\fc{r_B}{2})\rt)}\chi(u).
\eeq
Then, the perturbation equations can be rewritten as
\be
\chi''(u)=\lambda_0(u)\chi'(u)+s_0(u)\chi
\ee
where $\lambda_0$ and $s_0$ are functions of $u$ depending on the effective potential. The functions $\lambda_0$ and $s_0$ are complicated, so we do not show them explicitly. The first twenty modes for the scalar field, electromagnetic field, and gravitational field are shown in Fig.~\ref{qnmsvt}. The WKB method is powerful on solving frequencies of low overtone QNMs. We compare the results obtained by the AIM and by the WKB method in Tables~\ref{frequencyofscalar}, \ref{frequencyofvector}, \ref{frequencyoftensor} for the scalar field, electromagnetic field, and gravitational field, respectively. We find that, for low overtone QNMs, the results obtained through the AIM are in good agreement with that of obtained by the WKB method. When the multipole number $l$ increases, the real parts of the QNM frequencies change more apparently than the imaginary parts, which can be seen from Fig.~\ref{qnmsvt} and Tables~\ref{frequencyofscalar}, \ref{frequencyofvector}, \ref{frequencyoftensor}. Note that, $r_B=0.5r_S$ is equivalent to $Q_{\text{m}}=\fc{2\sqrt{2}}{3}\bar{M}$, where $\bar{M}\equiv\fc{\sqrt{3}\kappa_4}{8\pi}M$.

\begin{table}[!htb]
\begin{center}
\begin{tabular}{|c|c|c|c|}
\hline
~$l$~               &~$n$~&~~~~~AIM~~~~~                    &~~~~WKB    \\
\hline
    ~~                   & & Re($\omega r_S$)~~~~Im($\omega r_S$)& Re($\omega r_S$)~~~~Im($\omega r_S$)\\
\hline
    0                    &  0    &          ~~~0.220856~~~~-0.166592 ~~   &  ~~  0.219664~~~~-0.166443~~\\
                         &  1    &           ~~~0.186884~~~~-0.534457 ~~   &  ~~  0.193004~~~~-0.537457~~\\
\hline
    1                    &  0    &          ~~ 0.586476~~~~-0.158533 ~~   &  ~~  0.586679~~~~-0.158525~~\\
                         &  1    &          ~~ 0.554754~~~~-0.488295 ~~   &  ~~  0.556358~~~~-0.487138~~\\
\hline
    2                    &  0    &          ~~ 0.967669~~~~-0.157641 ~~   &  ~~  0.967666~~~~-0.157648~~\\
                         &  1    &          ~~ 0.946096~~~~-0.478040 ~~   &  ~~  0.946082~~~~-0.478076~~\\
\hline

\end{tabular}\\
\caption{Frequencies of low overtone QNMs for the scalar field. The parameter $r_B$ is set to $r_B=0.5r_S$.}
\label{frequencyofscalar}
\end{center}
\end{table}

\begin{table}[!htb]
\begin{center}
\begin{tabular}{|c|c|c|c|}
\hline
~$l$~                &~$n$~&~~~~~AIM~~~~~                    &~~~~WKB    \\
\hline
    ~~                   &  & Re($\omega r_S$)~~~~Im($\omega r_S$)& Re($\omega r_S$)~~~~Im($\omega r_S$)\\
\hline
    1                    &  0    &          ~~~0.513377~~~~-0.152855 ~~   &  ~~  0.513302~~~~-0.153142~~\\
                         &  1    &          ~~ 0.476438~~~~-0.474101 ~~   &  ~~  0.476701~~~~-0.537457~~\\
\hline
    2                    &  0    &          ~~ 0.924716~~~~-0.155637 ~~   &  ~~  0.924714~~~~-0.155649~~\\
                         &  1    &          ~~ 0.901934~~~~-0.472399 ~~   &  ~~  0.901941~~~~-0.472447~~\\
\hline
    3                    &  0    &          ~~ 1.32049~~~~~-0.156374 ~~    &  ~~  1.32049~~~~-0.156375~~\\
                         &  1    &          ~~ 1.30408~~~~~-0.471908 ~~    &  ~~  1.30408~~~~-0.471914~~\\
\hline

\end{tabular}\\
\caption{Frequencies of low overtone QNMs for the electromagnetic field. The parameter $r_B$ is set to $r_B=0.5r_S$.}
\label{frequencyofvector}
\end{center}
\end{table}

\begin{table}[!htb]
\begin{center}
\begin{tabular}{|c|c|c|c|}
\hline
~$l$~               &~$n$~&~~~~~AIM~~~~~                    &~~~~WKB    \\
\hline
    ~~                   &  & Re($\omega r_S$)~~~~Im($\omega r_S$)& Re($\omega r_S$)~~~~Im($\omega r_S$)\\
\hline
    2                    &  0    &           ~~0.810272~~~~-0.147554  ~~   &  ~~  0.810467~~~~-0.147398~~\\
                         &  1    &          ~~0.785979~~~~-0.449209 ~~    & ~~   0.787063~~~~-0.447841~~\\
\hline
    3                    &  0    &          ~~ 1.24218~~~~-0.152877 ~~    &  ~~ 1.24220~~~~-0.152875~~\\
                         &  1    &          ~~ 1.22506~~~~-0.461665 ~~    & ~~ 1.22529~~~~-0.461580~~\\
\hline
    4                    &  0    &          ~~ 1.65149~~~~-0.154694 ~~    & ~~1.65149~~~~-0.154694~~\\
                         &  1    &          ~~ 1.63833~~~~-0.465851 ~~    & ~~1.63831~~~~-0.465856~~\\
\hline

\end{tabular}\\
\caption{Frequencies of low overtone QNMs for the gravitational field. The parameter $r_B$ is set to $r_B=0.5r_S$.}
\label{frequencyoftensor}
\end{center}
\end{table}

\begin{figure*}[htb]
\begin{center}
\subfigure[$ $]  {\label{qnmrbscalarRe}
\includegraphics[width=5.7cm]{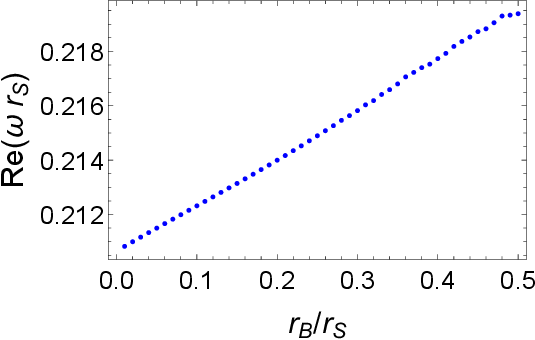}}
\subfigure[$ $]  {\label{qnmrbscalarIm}
\includegraphics[width=5.7cm]{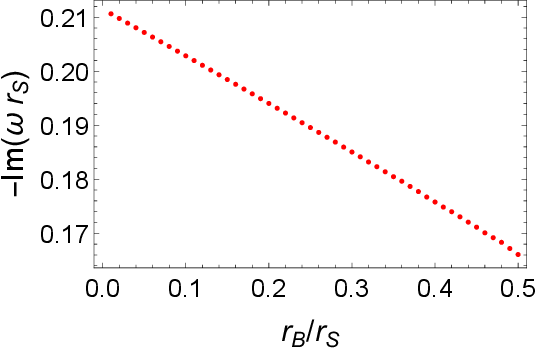}}
\subfigure[$ $]  {\label{qnmrbvectorRe}
\includegraphics[width=5.7cm]{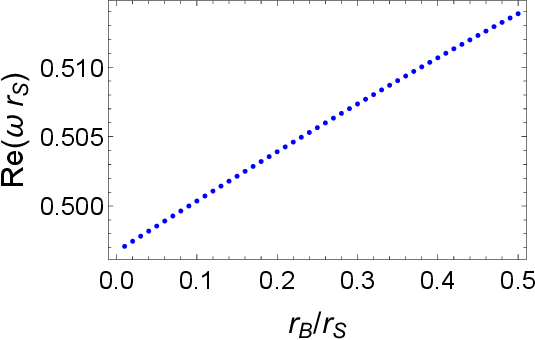}}
\subfigure[$ $]  {\label{qnmrbvectorIm}
\includegraphics[width=5.7cm]{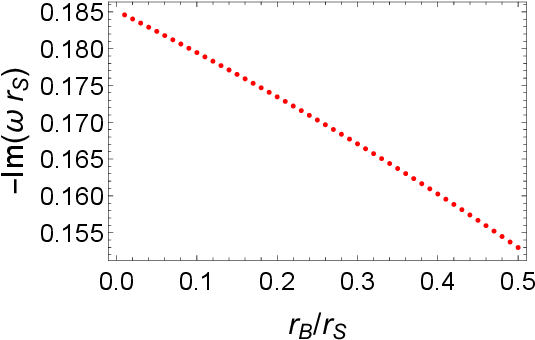}}
\subfigure[$ $]  {\label{qnmrbtensorRe}
\includegraphics[width=5.7cm]{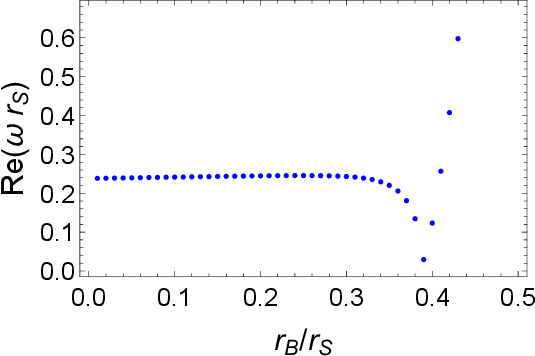}}
\subfigure[$ $]  {\label{qnmrbtensorIm}
\includegraphics[width=5.7cm]{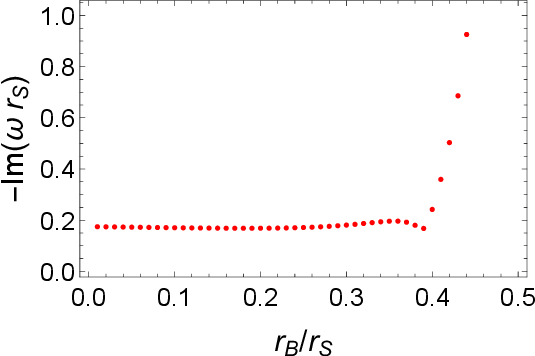}}
\end{center}
\caption{The effect of the parameter $r_B$ on the frequencies of fundamental QNMs.  (a) Real parts of frequencies for the scalar field with $l=0$. (b) Imaginary parts of frequencies for the scalar field with $l=0$. (c) Real parts of frequencies for the electromagnetic field with $l=1$. (d) Imaginary parts of frequencies for the electromagnetic field with $l=1$. (e) Real parts of frequencies for the gravitational field with $l=2$. (f) Imaginary parts of frequencies for the gravitational field with $l=2$.}
\label{qnmrbsvt}
\end{figure*}

From Eq.~\eqref{MQ2} we know that, the mass $M$ and the magnetic charge $Q_{\text{m}}$ are closely related to the parameter $r_B$ and $r_S$. When we fix the mass $M$ and increase the magnetic charge $Q_{\text{m}}$, then $r_S$ will decrease and $r_B$ will increase. So the effect of the magnetic charge $Q_{\text{m}}$ on the QNMs can be obtained through qualitatively analyse the effect of the parameter $r_B$ on the QNMs. We study the effect of the parameter $r_B$ on the fundamental QNMs for the three kinds of perturbed fields. The range of the parameter $r_B$ is $0\leq r_B\leq 0.5 r_S$. It is equivalent that $0\leq Q_{\text{m}}\leq \fc{2\sqrt{2}}{3}\bar{M}$. We find that the real parts of the overtone QNMs' frequencies for scalar field and vector field approximately increase linearly with $r_B$, while the absolute value of the imaginary parts approximately decrease linearly with $r_B$, which can be seen from Figs.~\ref{qnmrbscalarRe}-\ref{qnmrbvectorIm}. As for the gravitational field, both the real parts and the absolute value of the imaginary parts of the overtone QNM frequencies change slightly with $r_B$ for smaller $r_B$, after a short decrease stage, then they increase rapidly with $r_B$.

\subsection{Time evolution}

\begin{figure*}[htb]
\begin{center}
\subfigure[ $ $  ]  {\label{TDscalar}
\includegraphics[width=5.7cm]{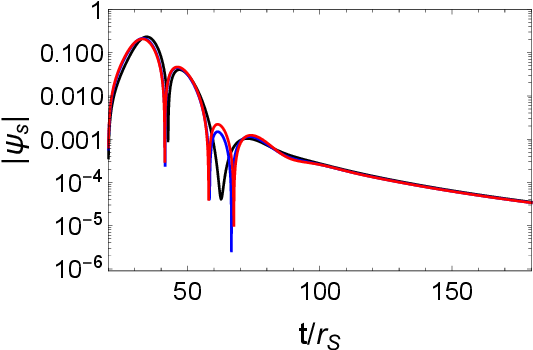}}
\subfigure[$ $]  {\label{TDvector}
\includegraphics[width=5.7cm]{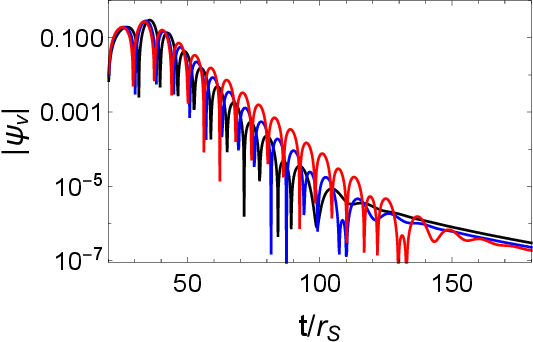}}
\subfigure[$ $]  {\label{TDtensor}
\includegraphics[width=5.7cm]{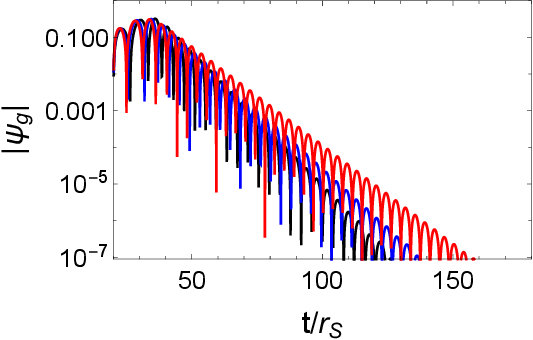}}
\end{center}
\caption{Time evolution of the Gauss package extracted at $r_*=20$. The parameter $r_B$ is set as $r_B=0.2r_S$ (the black lines), $r_B=0.5r_S$ (the blue lines), and $r_B=0.8r_S$ (the red lines). (a) Time evolution of the scalar field with $l=0$. (b)  Time evolution of the electromagnetic field with $l=1$. (c)  Time evolution of the gravitational field with $l=2$.}
\label{TDsvt}
\end{figure*}

In order to intuitively show the evolution of the perturbed field, we study the QNMs in the time domain, i.e., do not use the  ansatz $\psi\propto e^{-i \omega t}$. Using the null coordinates $u=t-r_*$ and $v=t+r_*$, the perturbation equations can be written in the following form
\be
4\fc{\pd^2\Phi_{s,v,g}}{\pd u\pd v}+V_{s,v,g}(r_*)\Phi_{s,v,g}=0.
\ee
We choose the initial data as a Gaussian package
\beq
\Phi(0,v)&=&e^{-\fc{(v-v_c)^2}{2\sigma^2}},\nn\\
\Phi(u,0)&=&0.
\eeq
We choose the package located at $v_c=10r_S$, with the width $\sigma=1r_S$. The evolution ranges are $(0,200r_S)$, and the results are extracted at $r_*=20r_S$. The results are shown in Fig.~\ref{TDsvt}. By fitting the evolution data, we can also obtain the QNM's frequency. For example, the frequency by fitting the evolution data of the electromagentic field is $0.512894-0.152494i$, which agrees with the result by the AIM $0.513377-0.152855i$ well. Although the fundamental QNM dominates the evolution of the perturbation, the evolution data are the superpositions of all the QNMs, so this
result is good. All three methods obtain the same results enhance the credibility of the results.

\section{Conclusions}\label{conclusion}

Through the KK reduction, the five-dimensional Einstein-Maxwell theory reduces to a four-dimensional Einstein-Maxwell-dilaton theory which supports a spherically static charged black hole solution. We studied the linear perturbation equations of scalar field, electromagnetic field, and gravitational field on this spherically static charged black hole background. Because of the spherical symmetry of the background, the radial parts of the perturbed fields can be decomposed from the angular parts. Using the tortoise coordinate $r_*$, every perturbation equation can written into a Schr\"{o}dinger-like form. The effective potentials for the scalar field, electromagnetic field, and gravitational field are shown in Fig.~\ref{potential}. From this figure we can see that, the effective potentials, except for that of the electromagnetic field, depend on  the parameter $r_B$.

Using the AIM we computed the QNM frequencies for the three kinds of perturbed fields. As the multipole number $l$ increases, the real parts of the QNM frequencies change more apparently than the imaginary parts, which can be seen from Fig.~\ref{qnmsvt} and Tables~\ref{frequencyofscalar}, \ref{frequencyofvector}, \ref{frequencyoftensor}. We also compared the results obtained from the AIM and WKB method, and found that, for low overtone QNMs, the results obtained through AIM are in good agreement with that of obtained by WKB method. The effect of the parameter $r_B$ were also studied. Using the null coordinates $u$ and $v$, the evolution of a Gaussian package was also investigated. The results showed that, the QNM frequencies obtained by fitting the evolution data agree well with the results by the AIM.

Note that, we only studied the QNMs for the charged black hole. For the topological star, there is no event horizon, so the ingoing boundary condition can not be imposed. Thus, it should be treated separately. We will study this in the future.

\section{Acknowledgments}

This work was supported by National Key Research and Development Program of China (Grant No. 2020YFC2201503), the National Natural Science Foundation of China (Grants No. 12205129, No. 12147166, No. 11875151, No. 12075103, and No. 12247101), the China Postdoctoral Science Foundation (Grant No. 2021M701529), the 111 Project (Grant No. B20063), and Lanzhou City's scientific research funding subsidy to Lanzhou University.

%

%
%

\end{document}